Nonreciprocal spin-charge interconversion in topological insulator/ferromagnet heterostructures

Shuyuan Shi[1,2,3,*], Enlong Liu[1], Fanrui Hu[1], Guoyi Shi[1], Aurélien Manchon[4,*], and Hyunsoo Yang[1,*]

[1]Department of Electrical and Computer Engineering, National University of Singapore, 117583, Singapore.
[2]Fert Beijing Institute, MIIT Key Laboratory of Spintronics, School of Integrated Circuit Science and Engineering, Beihang University, Beijing 100191, China.
[3]National Key Laboratory of Spintronics, Hangzhou International Innovation Institute, Beihang University, Hangzhou 311115, China.
[4]Aix-Marseille Université, CNRS, CINaM, Marseille, France.

*smeshis@buaa.edu.cn, aurelien.MANCHON@univ-amu.fr, eleyang@nus.edu.sg

**The process of spin-charge interconversion is critical in modern spintronics. Nonetheless, experiments conducted on a wide variety of magnetic heterostructures consistently report that charge-to-spin and spin-to-charge conversion efficiencies can be vastly different, especially in the case of topological insulators (TI). This discrepancy between the two "reciprocal" effects remains unexplained, hampering the development of spin devices based on spin-charge conversion. In this study, we investigate both spin-charge and charge-spin interconversion processes in TI $Bi_2Te_3$/Py and Pt/Py bilayers experimentally using spin-torque ferromagnetic resonance and spin pumping techniques. We find that the measured charge-to-spin conversion efficiency ($\xi_{C-S}$) in TI/Py is ~26 times larger than the measured spin-to-charge conversion efficiency ($\xi_{S-C}$), whereas $\xi_{C-S}$ and $\xi_{S-C}$ are comparable in the case of Pt/Py. Using a theoretical model enforcing Onsager reciprocity, we show that spin-to-**

**charge and charge-to-spin conversions in bilayers are genuinely inequivalent, and explain our results as arising from the distinct spin current leakage that takes place during the interconversion. This work clarifies previous conflicting reports on spin-charge interconversion processes and highlights the potential of interface engineering to achieve efficient spin transport in TI-based ferromagnetic heterostructures, paving the way for highly efficient spintronic devices.**

## I. INTRODUCTION

The charge-to-spin and spin-to-charge conversion processes in topological insulators (TIs) have garnered significant attention due to their potential for efficient spin current generation, detection, spin manipulation [1-6]. So far, much emphasis has been placed on the charge-to-spin conversion for efficient spin-orbit torque applications. On the other hand, novel device concepts utilizing spin-to-charge conversion for energy efficient beyond-CMOS computing have been proposed [7,8]. Because of these new opportunities, an accurate evaluation of spin-to-charge conversion is crucial for understanding and effectively utilizing these devices. Despite numerous experiments conducted to assess $\xi_{C-S}$ and $\xi_{S-C}$ in TIs, there remains substantial controversy over the reported values. Notably, extremely large $\xi_{C-S}$ values, ranging from 0.4 to 425, have been observed in various TIs such as $Bi_2Se_3$, $Bi_2Te_3$, $Bi_2Sb_3$, and $(Bi_{0.5}Sb_{0.5})_2Te_3$ [1,2,4,6,9]. However, previous reports indicate that $\xi_{S-C}$ values are generally smaller than $\xi_{C-S}$, ranging from 0.0001 to 0.4 [3,5,10-12]. Given that the charge-to-spin conversion process and its inverse are theoretically reciprocal [13], the exceedingly small $\xi_{S-C}$ value measured in TIs using spin pumping was previously attributed to a relatively poor TI/FM interface [3]. In fact, spin-charge interconversion efficiencies are influenced by factors such as different TI materials, TI thicknesses, carrier concentrations, measurement temperatures, and interface spin transparency, all of which vary across different studies [2,4,14,15]. Therefore, it is essential to evaluate both $\xi_{C\text{-}S}$ and $\xi_{S\text{-}C}$ within the same material system and under consistent measurement conditions.

In this work, we investigate $\xi_{C\text{-}S}$ and $\xi_{S\text{-}C}$ using spin-torque ferromagnetic resonance (ST-FMR) and spin pumping (SP) techniques, respectively, in the same $Bi_2Te_3$ (13 QL)/$Ni_{79}Fe_{21}$ (Py) bilayer stack at room temperature. We observe significantly different values for $\xi_{C\text{-}S}$ and $\xi_{S\text{-}C}$ in $Bi_2Te_3$/Py, while our control experiments with Pt/Py reveal comparable $\xi_{C\text{-}S}$ and $\xi_{S\text{-}C}$ values. By maintaining a consistent TI/FM interface, we can rule out interface conditions as the primary cause of the differing $\xi_{C\text{-}S}$ and $\xi_{S\text{-}C}$ values in $Bi_2Te_3$/Py. Instead, the one order-of-magnitude difference in spin transmission between the topological surface states (TSS) of the TI and the adjacent Py layer results in substantial spin leakage during the spin-to-charge conversion process, leading to a drastic reduction of $\xi_{S\text{-}C}$. By inserting a 0.8 nm thick-MgO layer between $Bi_2Te_3$ and Py, the difference between $\xi_{C\text{-}S}$ and $\xi_{S\text{-}C}$ in $Bi_2Te_3$/MgO/Py is found to be smaller than that in $Bi_2Te_3$/Py due to longer effective spin relaxation, which supports our model.

**II. EXPERIMENT**

$Bi_2Te_3$ films of 13 quintuple layers (QL, 1 QL ~ 1 nm) were grown on $Al_2O_3$ (0001) substrates using molecular beam epitaxy (MBE). The growth method is detailed in Supplemental Material Sec. A [16] (see also references [17-22] therein). Figure 1(a) shows an atomic force microscopy (AFM) image of a representative 13-QL $Bi_2Te_3$ film, revealing a smooth surface with a roughness of 0.5 nm. The Raman spectrum in Fig. 1(b) displays the resonance modes of $A_{1g}^1$, $E_g^2$ and $A_{1g}^2$ of the $Bi_2Te_3$ film, consistent with previous reports [23,24]. An X-ray diffraction (XRD) scan of the 13 QL $Bi_2Te_3$ film, shown in Fig. 1(c), indicates a phase-pure $Bi_2Te_3$ layer. Four-probe measurements show that the resistivity ($\rho_{BiTe}$) is ~1470 µΩ·cm at room temperature, aligning with a previous report [25]. The temperature-dependent resistivity in Fig. 1(d) demonstrates the typical metallic behavior in 13 QL $Bi_2Te_3$ films [26].

We prepared $Bi_2Te_3$ (13 QL)/Py (6 nm)/$SiO_2$ (5 nm) heterostructures by depositing 6 nm of Py and 5 nm of $SiO_2$ on top of $Bi_2Te_3$ using sputtering with a base pressure $< 2\times10^{-9}$ Torr. The 5 nm-$SiO_2$ layer was deposited to protect the $Bi_2Te_3$ (13 QL)/Py (6 nm) bilayer from oxidation. Using photolithography and Ar-ion milling, we fabricated the heterostructures into ST-FMR and SP devices of varying sizes to study the charge-to-spin and spin-to-charge conversion processes, respectively. Details on the device fabrication are provided in Supplemental Material Sec. B [16]. For reference, we also fabricated ST-FMR and SP devices with heavy metal Pt (5 nm)/Py (6 nm)/$SiO_2$ (5 nm). Figures 2(a) and (b) illustrate the configurations for the ST-FMR and SP measurements, respectively. Both measurements used radio frequency (r.f.) input signals ranging from 6 to 10 GHz. All measurements were conducted at room temperature.

### III. RESULTS AND DISCUSSION

During the ST-FMR measurements, an r.f. current ($I_{RF}$; current density $J_C$ in the $Bi_2Te_3$ layer) with a power of 13 dBm was applied to the device. The generated spin currents $J_S$, indicated by red arrows in Fig. 2(a), diffuse into the Py layer, exerting spin-orbit torques (SOT) on the Py layer. Consequently, the Py magnetization is excited into the precession mode, generating an ST-FMR voltage $V_{mix}$, as shown in Fig. 2(c) for a $Bi_2Te_3$/Py ST-FMR device. In Fig. 2(d), the $V_{mix}$ (open symbols) at a frequency of 6 GHz is fitted by $V_{mix} = V_S F_S + V_A F_A$, where $V_S F_S$ and $V_A F_A$ are the symmetric and antisymmetric Lorentzian components, respectively. Using an established analysis method described in Supplemental Material Sec. C [16], we determine the charge-to-spin conversion efficiency ($\xi_{C-S} = J_S/J_C$) based solely on the symmetric component $V_S F_S$ for both $Bi_2Te_3$ and Pt devices. The light blue columns on the left and right in Fig. 3(a) show the $\xi_{C-S}$ values for $Bi_2Te_3$/Py and Pt/Py, respectively. The error bars from independent measurements on three different devices. The averaged $\xi_{C-S}$ values for $Bi_2Te_3$/Py and Pt/Py are 0.52 ± 0.01 and 0.063 ± 0.001, respectively. From the ST-FMR measurements, $\xi_{C-S}$ in $Bi_2Te_3$ is about an order of magnitude larger than in Pt at room temperature, consistent with previous studies [25,27]. The angular-dependent ST-FMR measurements have also been performed on $Bi_2Te_3$/Py and Pt/Py samples to exclude the artifact signal (Supplemental Fig. S1) [16,28].

Next, we investigate the spin-to-charge conversion process in both $Bi_2Te_3$ and Pt using the SP technique. The fabrication process of the SP devices is detailed in Supplemental Material Sec. B [16]. In the SP measurements, as illustrated in Fig. 2(b), an external d.c. magnetic field $H$ is swept along the y-axis. An r.f. current is applied to the waveguide, inducing an r.f. magnetic field $h_{rf}$ in the Py layer, leading to ferromagnetic resonance (FMR). As a consequence, a spin current density $J_S$ is generated in the Py, which then diffuses across the $Bi_2Te_3$/Py interface into the $Bi_2Te_3$ layer. The spin current is subsequently converted into a transverse charge current with a density $J_C$, resulting in a SP voltage $V_{mix}$, which is measured using a lock-in amplifier. Figure 2(e) shows the SP signal $V_{mix}$ from a $Bi_2Te_3$/Py device at various frequencies. Similar to ST-FMR, $V_{mix}$ in SP can be fitted with $V_{mix} = V_S F_S + V_A F_A$ as shown in Fig. 2(f). We extract the spin-to-charge conversion efficiency ($\xi_{S-C} = J_C/J_S$) for both $Bi_2Te_3$ and Pt SP devices using the method detailed in Supplemental Material Sec. D [16]. The orange columns on the left and right in Fig. 3(a) represent $\xi_{S-C}$ in $Bi_2Te_3$/Py and Pt/Py, respectively, with error bars indicating the standard deviations of $\xi_{S-C}$ from three different SP devices. The averaged $\xi_{S-C}$ values for $Bi_2Te_3$/Py and Pt/Py are 0.020 ± 0.007 and 0.034 ± 0.006, respectively. Notably, the measured $\xi_{S-C}$ of $Bi_2Te_3$ is even smaller than that of Pt, which contrasts with the ST-FMR results, where $\xi_{C-S}$ is 26 times larger than $\xi_{S-C}$ in Bi$_2$Te$_3$/Py. However, this finding is qualitatively consistent with previous SP measurements [3,5,10-12].

To understand the large difference between $\xi_{S-C}$ and $\xi_{C-S}$ in Bi$_2$Te$_3$/Py, but not in Pt/Py, we first examine the Gilbert damping coefficient ($\alpha$) and the real part of the effective spin mixing conductance ($g_{\text{eff}}^{\uparrow\downarrow}$, equivalently $G_{\text{eff}}^{\uparrow\downarrow} = \frac{e^2}{h} g_{\text{eff}}^{\uparrow\downarrow}$,). The coefficient $\alpha$ is determined from the relation $\Delta = \Delta_0 + 2\pi\alpha f/\gamma$ [29,30], where $\Delta$ the linewidth, $\Delta_0$ is the inhomogeneous linewidth broadening, and $\gamma$ is the gyromagnetic ratio. We calculate $\alpha$ based on ST-FMR and SP measurements. Figure 3(b) shows that $\alpha$ is similar within each material group (Bi$_2$Te$_3$/Py and Pt/Py), regardless of measurement techniques such as ST-FMR and SP. The difference in α between Bi$_2$Te$_3$/Py and Pt/Py is attributed to the different material interfaces [31]. The average α values for Bi$_2$Te$_3$/Py and Pt/Py are 0.036 and 0.022, respectively, consistent with previous reports [10,32].

Next, we extract $g_{\text{eff}}^{\uparrow\downarrow}$ from both ST-FMR and SP devices for the Bi$_2$Te$_3$/Py and Pt/Py heterostructures. The value of $g_{\text{eff}}^{\uparrow\downarrow}$ at the non-magnet/ferromagnet interface characterizes the interfacial spin transport efficiency and can be determined using $g_{\text{eff}}^{\uparrow\downarrow} = \frac{4\pi M_s t_{\text{FM}}}{g\mu_B}(\alpha - \alpha_0)$ [31,33], where $\alpha_0$ is the thickness-independent intrinsic damping of Py layer, $M_s$ is the effective magnetization of the Py layer, $t_{\text{FM}}$ is the thickness of the Py layer, $g$ is the Landé factor and $\mu_B$ is the Bohr magnetron. As shown in Fig. 3(c), similar values of $g_{\text{eff}}^{\uparrow\downarrow}$ are obtained for each Bi$_2$Te$_3$/Py and Pt/Py heterostructure, regardless of the measurement technique (ST-FMR versus SP). The comparable $\alpha$ and $g_{\text{eff}}^{\uparrow\downarrow}$ values in the Bi$_2$Te$_3$/Py and Pt/Py heterostructures suggest that the charge-to-spin (ST-FMR) and spin-to-charge (SP) conversion processes occur under similar interfacial conditions. Therefore, differences in the interfacial quality do not account for the disparity between $\xi_{C-S}$ and $\xi_{S-C}$ in Bi$_2$Te$_3$/Py measured using ST-FMR and SP, respectively. In addition, the discrepancy in $\xi_{C-S}$ and $\xi_{S-C}$ in Bi$_2$Te$_3$/Py cannot be attributed to the type of TI material, TI thickness, TI quality, nor measurement temperature, as these variables are consistent across our ST-FMR and SP devices.

To explain the observed differences between Pt/Py and Bi$_2$Te$_3$/Py, we exploit the model introduced in Ref. [34]. In a nutshell, this model describes the spin-charge interconversion in a magnetic bilayer in the presence of both the spin Hall effect (SHE) in the normal metal and the spin Rashba-Edelstein effect (REE) at the interface, as depicted in Figs. 4(a) and (b). To account for both effects, the nonmagnetic layer is characterized by its thickness $d_N$, conductivity $\sigma_N$, spin relaxation length $\lambda_N$, and spin Hall angle $\theta_H$, whereas the interface is modeled as an effective layer of thickness $d_i$, spin relaxation length $\lambda_i$, and the effective Rashba parameter $\alpha_R$. The coupling between the ferromagnetic layer and the interface is governed by the spin-mixing conductance $G_{\text{eff}}^{\uparrow\downarrow}$, whereas the coupling between the interface and the normal metal is controlled by the interfacial conductance $G_N$. Finally, the lifetime of the spin in the interfacial layer is modelled by the conductance $G_i$ (see below). With this parameterization, the charge-to-spin and spin-to-charge interconversion efficiencies read [34]

$$\xi_{C-S} = 2G_{\text{eff}}^{\uparrow\downarrow} \frac{d_F + d_N + d_i}{\sigma_N d_N + \sigma_i} \frac{e^2 \mathcal{N}_i \frac{\alpha_R}{\hbar} + \tilde{\theta}_H \tilde{\lambda}_N \frac{G_N}{1+\eta_N}}{G_i + 2G_{\text{eff}}^{\uparrow\downarrow} + \frac{G_N}{1+\eta_N}}, \qquad (1)$$

$$\xi_{S-C} = \frac{1}{d_N + d_i} \frac{e^2 \mathcal{N}_i \frac{\alpha_R}{\hbar} + \tilde{\theta}_H \tilde{\lambda}_N \frac{G_N}{1+\eta_N}}{G_i + \frac{G_N}{1+\eta_N}}. \qquad (2)$$

In these expressions, $d_F$ is the thickness of the ferromagnet, $\mathcal{N}_i$ is the density of states of the interface, and we define the effective spin Hall angle $\tilde{\theta}_H = \theta_H \left(1 - \cosh \frac{d_N}{\lambda_N}\right)^{-1}$, the effective spin relaxation length $\tilde{\lambda}_N = \lambda_N / \tanh \frac{d_N}{\lambda_N}$, and the spin transparency $1/(1+\eta_N)$, with $\eta_N = \tilde{\lambda}_N G_N / \sigma_N$.

The spin transparency accounts for the ability of the nonmagnetic metal to absorb an incoming spin current: the larger the conductivity and the smaller the spin relaxation length, the larger the spin transparency. Whereas the spin-charge and charge-spin interconversions are driven by the same mechanisms (REE and SHE), $\sim e^2 \mathcal{N}_i \frac{\alpha_R}{\hbar} + \tilde{\theta}_H \tilde{\lambda}_N \frac{G_N}{1+\eta_N}$, the conversion efficiencies differ in two ways. First, because of the different current distributions of these two processes (see Ref. [34]), they display markedly different geometrical factors, $\xi_{C-S} \sim \frac{d_F+d_N+d_i}{\sigma_N d_N+\sigma_i}$, $\xi_{S-C} \sim \frac{1}{d_N+d_i}$. In addition, the interplay between spin relaxation at the interface ($G_i$), tunnelling to/from the ferromagnet ($2G_{\uparrow\downarrow}$), and the absorption in the nonmagnetic metal ($\frac{G_N}{1+\eta_N}$) is different for these two processes, resulting in distinct dependences $\xi_{C-S} \sim \frac{2G_{\uparrow\downarrow}}{G_i+2G_{\uparrow\downarrow}+\frac{G_N}{1+\eta_N}}$, $\xi_{S-C} \sim \frac{1}{G_i+\frac{G_N}{1+\eta_N}}$. In other words, the interconversion efficiencies also depend on whether the nonmagnetic metal is a spin sink, $\frac{1}{1+\eta_N} \to 1$, or a spin insulator, $\frac{1}{1+\eta_N} \to 0$.

To fit the experimental data, we first fix the material parameters to their known values in Table I. We then set the interconversion coefficients to experimental values, i.e., $\xi_{C \to S} = 0.5, \xi_{S \to C} = 0.02$ in the case of Bi$_2$Te$_3$/Py, which places a constraint on the relation between $G_{\text{eff}}^{\uparrow\downarrow}$ and $G_i + \frac{G_N}{1+\eta_N}$. For instance, in the case of Bi$_2$Te$_3$/Py, we obtain

$$G_i + \frac{G_N}{1+\eta_N} = \frac{1.5 \times 10^{14}}{\left(1-\frac{1.5 \times 10^{14}}{2G_{\text{eff}}^{\uparrow\downarrow}}\right)} \approx 2.6 \times 10^{14} \, \Omega^{-1} \text{m}^{-2}. \tag{3}$$

Assuming $G_{\text{eff}}^{\uparrow\downarrow} = 2.8 \times 10^{15} \, \Omega^{-1} \text{m}^{-2}$ (Fig. 3c) and $G_N = 10^{14} \, \Omega^{-1} \text{m}^{-2}$ (see, e.g., Ref. [35]), we obtain $G_i \approx 7 \times 10^{13} \, \Omega^{-1} \text{m}^{-2}$, corresponding to a spin-flip rate of $\frac{1}{\tau_{sf}} = \frac{G_i}{e^2 \mathcal{N}} = 2 \times 10^{13} \, s^{-1}$.

These parameters lead to the following relation

$$e^2 \mathcal{N}_i \frac{\alpha_R}{\hbar} + \tilde{\theta}_H \tilde{\lambda}_N \frac{G_N}{1+\eta_N} = 2.5 \times 10^4 \, \Omega^{-1} \cdot m^{-1}. \tag{4}$$

As long as the conductivity of Bi$_2$Te$_3$ is very small, the SHE is inefficient because $\tilde{\theta}_H \tilde{\lambda}_N \frac{G_N}{1+\eta_N} \ll 2.5 \times 10^4 \, \Omega^{-1} \cdot m^{-1}$, whatever $\theta_H$ may be. In other words, the signal comes mainly from the interface and $e^2 \mathcal{N}_i \frac{\alpha_R}{\hbar} \approx 2.5 \times 10^4 \, \Omega^{-1} \cdot m^{-1}$. Assuming that $\mathcal{N}_i \approx 2 \times 10^{19}$ eV$^{-1}$m$^{-2}$, then we obtain $\alpha_R = 5 \times 10^{-12}$ eV$\cdot m$. Since $\mathcal{N}_i$ and $\alpha_R$ are not independent parameters, we cannot determine them accurately. This procedure is also applied to Pt/Py and the extracted parameters are reported in Table I.

To assess the robustness of these estimates, we adopted the parameters extracted from the experiments (see Table I), and we have computed the charge-spin and spin-charge interconversion efficiencies, $\xi_{C-S}$ and $\xi_{S-C}$, as a function of the conductivity of the normal metal. The results are reported in Fig. 4(c) for Bi$_2$Te$_3$/Py and in Fig. 4(e) for Pt/Py. The solid lines with symbols represent the simulated spin-to-charge (red) and charge-to-spin (blue) efficiencies. The horizontal dashed lines show the efficiencies measured experimentally, and the vertical dashed line indicates the nominal conductivity of the spin source layer (Bi$_2$Te$_3$ or Pt). In the case of Bi$_2$Te$_3$/Py [Fig. 4(c)], as expected, the SHE has a negligible impact on the spin-charge interconversion as long as the conductivity is low (except for unrealistically large spin Hall angles), and the REE dominates [Fig. 4(d)]. Increasing the conductivity drastically reduces the charge-to-spin conversion, while the spin-to-charge conversion is only mildly affected, pointing out the effect of the geometrical factor in Eq. (1, 2). In the case of Pt/Py [Fig. 4(e)], we find that the interfacial REE has a strong impact on both spin-charge and charge-spin conversions as long as the conductivity is low, but is negligible once the system becomes metallic [Fig. 4(f)].

To further assess the validity of our model, we insert a 0.8 nm MgO layer between Bi$_2$Te$_3$ and Py. The MgO tunnel barrier is expected to substantially reduce the coupling between the ferromagnet and the interfacial layer; in other words, we expect a strong reduction of the mixing conductance $G_{\uparrow\downarrow}$. ST-FMR and SP measurements on Bi$_2$Te$_3$/MgO/Py yield average values of $\xi_{C-S}$ = 0.077 ± 0.008 and $\xi_{S-C}$ = 0.0067 ± 0.0003, as shown in Fig. 3(d), alongside the values obtained for Bi$_2$Te$_3$/Py. Although the absolute values of $\xi_{C-S}$ and $\xi_{S-C}$ decrease with the insertion of the 0.8 nm MgO layer, the ratio of $\xi_{C-S}/\xi_{S-C}$ also decreases, from 26 in Bi$_2$Te$_3$/Py to 11.5 in Bi$_2$Te$_3$/MgO/Py. The reduction in absolute spin-charge interconversion efficiencies is primarily due to decreased spin transmission across the MgO layer. Indeed, applying our model to this system (see Table 1) and assuming $\alpha_R = 5 \times 10^{-12}$ eV · m (same as in Bi$_2$Te$_3$/Py) and $G_{\text{eff}}^{\uparrow\downarrow} \approx 5.5 \times 10^{14} \Omega^{-1} \cdot m^{-2}$ as measured from our experiments, we find that the conductivity to the two-dimensional (2D) gas is an order of magnitude larger than in Bi$_2$Te$_3$/Py. Furthermore, since the mixing conductance is associated with an escape time $\tau_{\uparrow\downarrow} = e^2 \mathcal{N}_i / G_{\text{eff}}^{\uparrow\downarrow}$, this model aligns with observations of an extremely large inverse REE in SrTiO$_3$/AlO$_x$/Py due to the much longer spin escape time in AlO$_x$ compared to the relaxation time in the 2D electron gas [36] and a relatively weak inverse REE in Ag/Bi due to the short relaxation time in metals [37].

## IV. SUMMARY

We have revisited the charge-to-spin ($\xi_{C-S}$) and spin-to-charge ($\xi_{S-C}$) conversion processes in both TI-based (Bi$_2$Te$_3$/Py) and heavy-metal-based (Pt/Py) systems. Our findings reveal that the effective $\xi_{C-S}$ is significantly larger than $\xi_{S-C}$ in Bi$_2$Te$_3$/Py, while the values of $\xi_{C-S}$ and $\xi_{S-C}$ are comparable in the Pt/Py system. By analyzing the spin-charge interconversion processes using a model treating SHE and REE on equal footing, we attribute these differences to the combination of two factors, a geometrical factor that reflects the current distribution in the bilayer, and the spin absorption of the nonmagnetic metal that is much larger in TI than in Pt. This work enhances our understanding of spin relaxation at the TI/FM interface and its impact on accurately quantifying spin-charge interconversion efficiencies. Additionally, our results suggest a potential pathway for achieving more efficient TI-based spintronic devices such as memory device (charge-to-spin conversion), and magneto-electric spin-orbit (MESO) transistors (spin-to-charge conversion).

## ACKNOWLEDGMENTS

This work was supported by National Research Foundation (NRF) Singapore Investigatorship (NRFI06-2020-0015), MOE Tier 2 (T2EP50123-0035), and Samsung Electronics Co., Ltd (IO241218-11518-01). S.S. acknowledges the financial support by the National Natural Science Foundation of China (grant No. 62474015), the Beijing Nova Program (20240484654), and Beihang World TOP University Cooperation Program. A.M. acknowledges support by the ANR ORION project, grant ANR-20-CE30-0022-01 of the French Agence Nationale de la Recherche and by France 2030 government investment plan managed by the French National Research Agency under grant reference PEPR SPIN – [SPINTHEORY] ANR-22-EXSP-0009.

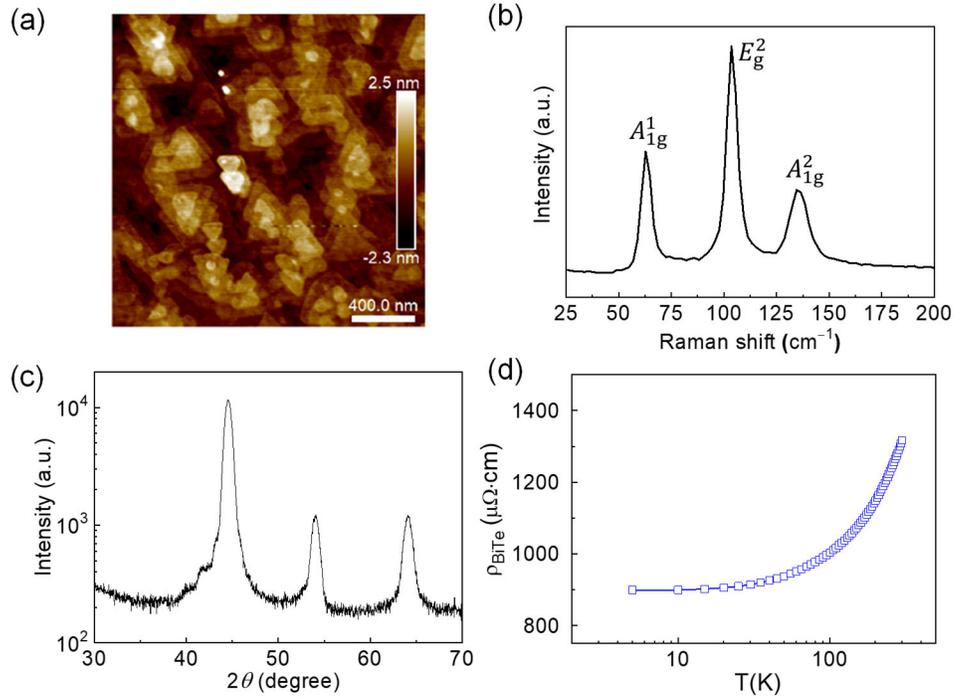

FIG. 1. (a) AFM image of a 13 QL Bi$_2$Te$_3$ film. The surface roughness is ~0.5 nm. (b) Raman spectrum of a 13 QL Bi$_2$Te$_3$ film, showing the representative resonance modes of $A_{1g}^1$, $E_g^2$ and $A_{1g}^2$ of the Bi$_2$Te$_3$ film. (c) Representative $\theta-2\theta$ x-ray diffraction (XRD) scan of a 13 QL Bi$_2$Te$_3$ film, indicating a phase-pure Bi$_2$Te$_3$ layer. (d) Temperature dependent resistivity of a 13 QL Bi$_2$Te$_3$ film, showing a typical metallic behavior.

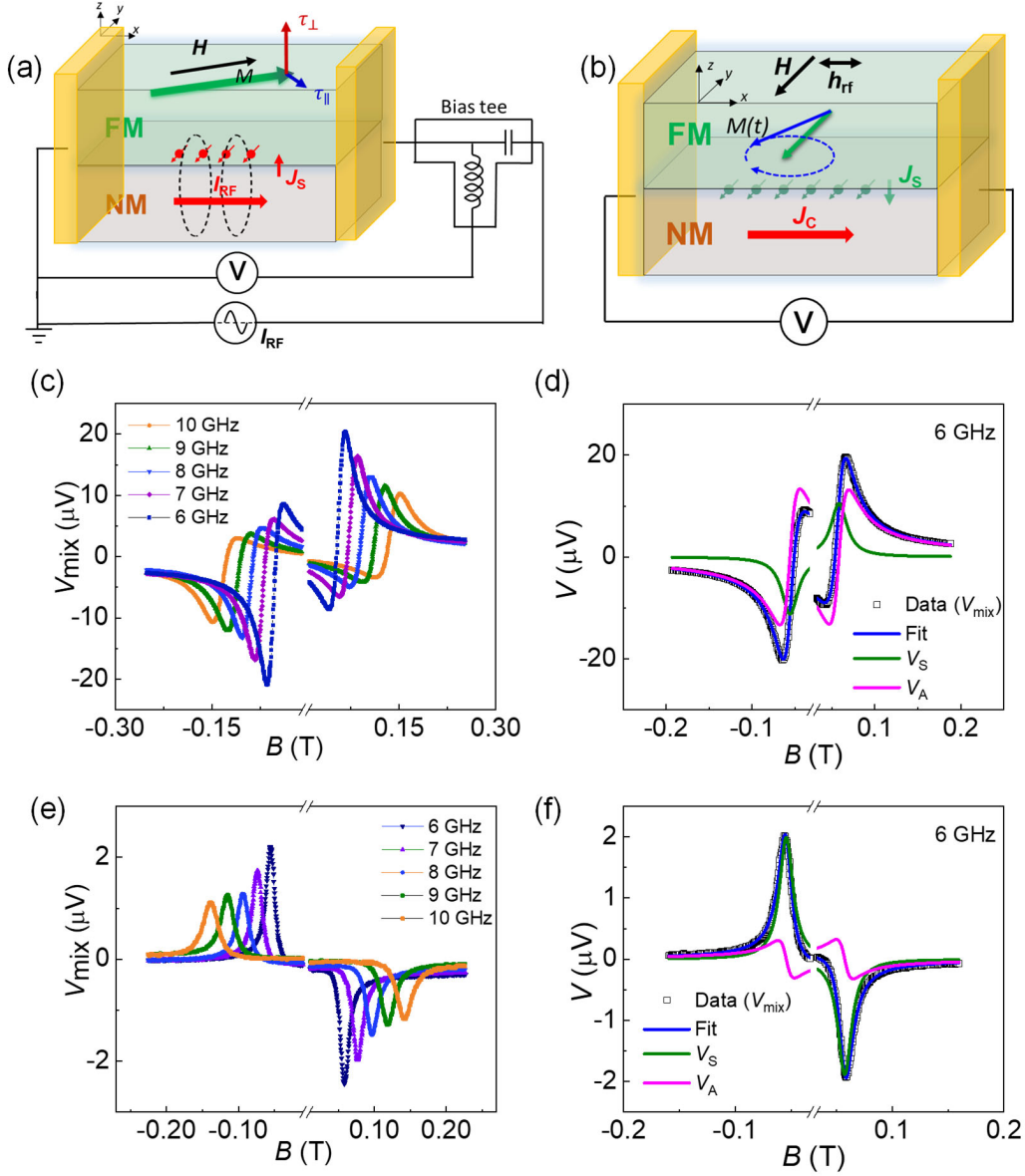

FIG. 2. (a) Schematic diagram of the ST-FMR experimental configuration which consists of a ferromagnet (FM) and a nonmagnetic layer (NM). $\tau_{\parallel}$ is the in-plane spin-orbit torque, and $\tau_{\perp}$ is the out-of-plane torque. $I_{RF}$ is the injected radio frequency (r.f.) current. The external magnetic field $H$ is applied with an angle $\theta$ with respect to $I_{RF}$. $J_S$ is the spin current density generated by NM. (b) Schematic diagram of the SP device and experimental configuration. M($t$) is the magnetization, and $H$ and $h_{rf}$ are d.c. and r.f. magnetic fields, respectively. $J_S$ is the spin current generated by the FM layer. $J_C$ is the charge current density converted from $J_S$. (c) Representative ST-FMR spectra for a $Bi_2Te_3$/Py device with the r.f. current frequency from 6 to 10 GHz. (d) ST-FMR signal of the $Bi_2Te_3$/Py sample at 6 GHz. The solid lines are fits showing the symmetric ($V_S$) (green) and antisymmetric ($V_A$) (magenta) Lorentzian contribution. (e) Representative SP spectra for a $Bi_2Te_3$/Py device with the r.f. current frequency from 6 to 10 GHz. (f) SP signal of the $Bi_2Te_3$/Py sample at 6 GHz. The solid lines are fits showing the symmetric ($V_S$) (green) and antisymmetric ($V_A$) (magenta) Lorentzian contributions.

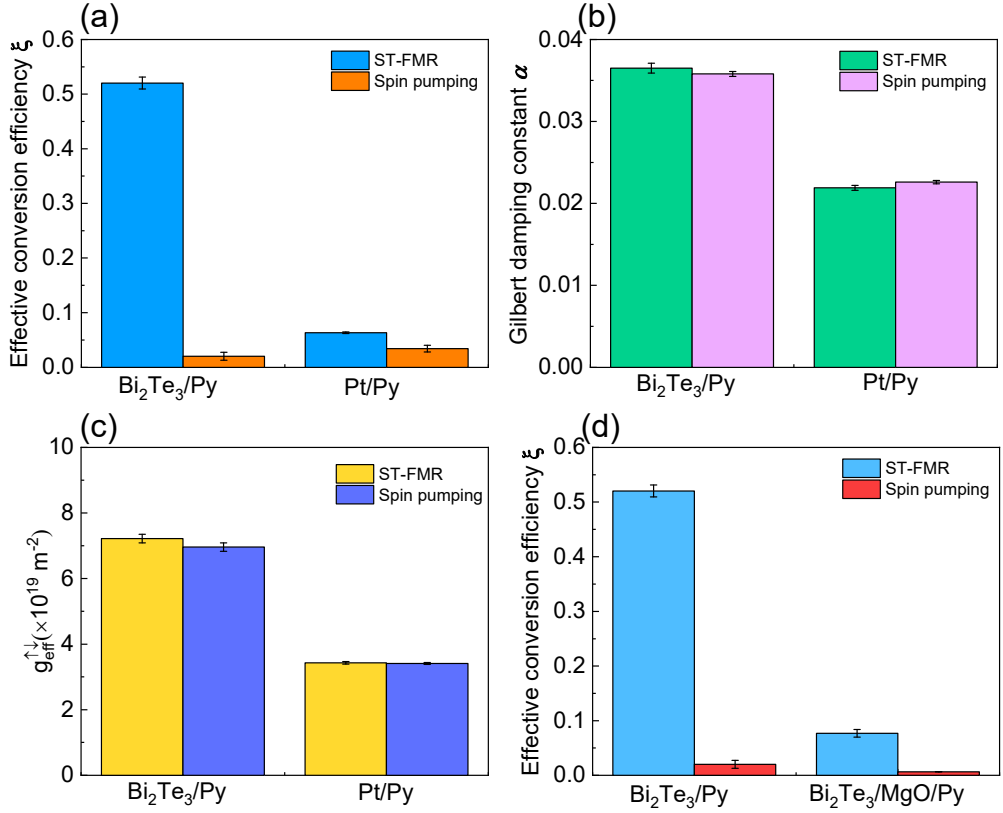

FIG. 3. (a) Effective charge-to-spin conversion efficiency $\xi_{C\text{-}S}$ (light blue) and spin-to-charge conversion efficiency $\xi_{S\text{-}C}$ (orange) measured respectively by ST-FMR and SP measurements in $Bi_2Te_3$/Py and Pt/Py. (b) Gilbert damping coefficient ($\alpha$) measured based on $Bi_2Te_3$/Py and Pt/Py heterostructures using ST-FMR (green) and SP (pink) measurements. (c) Effective spin mixing conductance ($g_{\text{eff}}^{\uparrow\downarrow}$) at the $Bi_2Te_3$/Py interface and the Pt/Py interface using ST-FMR (yellow) and SP (blue) measurements. (d) $\xi_{C\text{-}S}$ (light blue) and $\xi_{S\text{-}C}$ (red) measured respectively by ST-FMR and SP measurements in $Bi_2Te_3$/Py and $Bi_2Te_3$/MgO/Py.

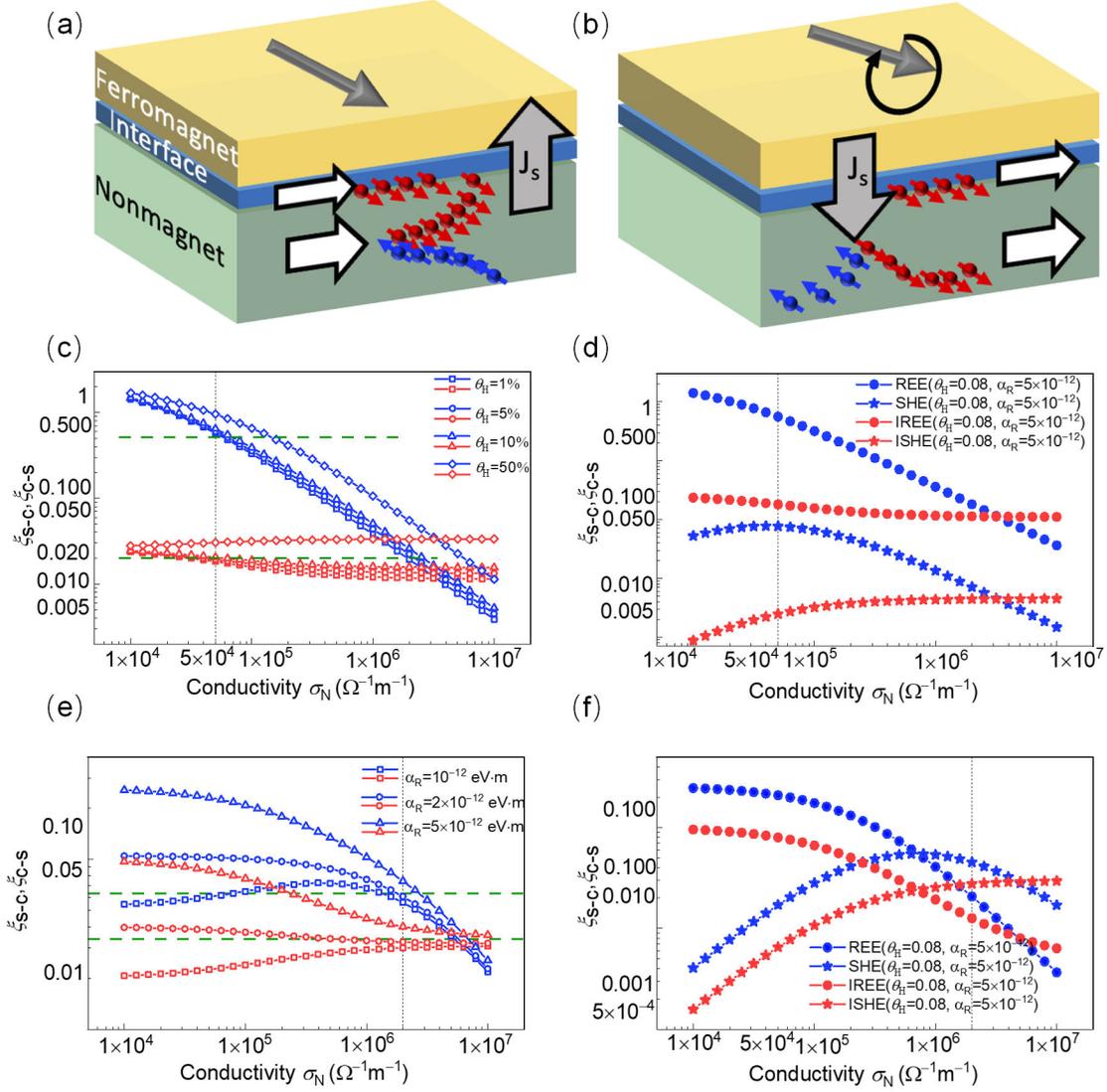

FIG. 4. Sketch of the (a) spin-to-charge and (b) charge-to-spin conversion processes accounting for both SHE/ISHE and IREE/REE. (c,e) charge-to-spin ($\xi_{c-s}$; blue symbol) and spin-charge ($\xi_{s-c}$; red symbol) conversion efficiencies as a function of the conductivity of the normal metal, for different spin Hall angles, using the parameters of (c) $Bi_2Te_3$/Py and of (e) Pt/Py, respectively. (d,f) Contribution of the inverse Rashba-Edelstein (circle symbol) and spin Hall effects (star symbol) for the charge-spin (blue) and spin-charge (red) efficiencies for (d) $Bi_2Te_3$/Py and (f) Pt/Py. The vertical dashed lines indicate the experimental conductivity of (c,d) $Bi_2Te_3$ and (e,f) Pt.

TABLE 1. Materials parameters extracted from the theoretical fit of the experimental results. The layer thickness, spin relaxation length, conductivity, spin Hall angle and the spin mixing conductances are fixed to known values. The other parameters are determined during the fitting procedure. For Bi$_2$Te$_3$/MgO/Py, we assumed that the values of $\sigma_N$, $G_N$ and $\alpha_R$ are the same as in Bi$_2$Te$_3$/Py (i.e., unaffected by MgO insertion). Note that because of the strong difference in conductivities between Bi$_2$Te$_3$ and Pt, the exact magnitude of $\theta_H$ ($\alpha_R$) is unimportant in the former (latter) since spin-charge interconversion is governed by REE (SHE).

| Fitting parameters | Bi$_2$Te$_3$/Py | Bi$_2$Te$_3$/MgO/Py | Pt/Py with Rashba surface |
|---|---|---|---|
| $d_F$ | | 6 nm | |
| $d_N$ | 12 QL | 12 QL | 4.5 nm |
| $d_{2D}$ | 1 QL | 1 QL | 0.5 nm |
| $\sigma_N$ | $7 \times 10^4 \, \Omega^{-1} \cdot m^{-1}$ | $7 \times 10^4 \, \Omega^{-1} \cdot m^{-1}$ | $2 \times 10^6 \, \Omega^{-1} \cdot m^{-1}$ |
| $\lambda_N$ | 1 nm | 1 nm | 3 nm |
| $\theta_H$ | < 10% | < 10% | 8% |
| $\sigma_{2D}$ | $2 \times 10^{-4} \, \Omega^{-1}$ | $4.5 \times 10^{-3} \, \Omega^{-1}$ | $\sigma_N \times d_{2D}$ |
| $G_{eff}^{\uparrow\downarrow}$ | $2.8 \times 10^{15} \Omega^{-1} \cdot m^{-2}$ | $5.5 \times 10^{14} \Omega^{-1} \cdot m^{-2}$ | $1.3 \times 10^{15} \Omega^{-1} \cdot m^{-2}$ |
| $G_N$ | $10^{14} \Omega^{-1} \cdot m^{-2}$ | $10^{14} \Omega^{-1} \cdot m^{-2}$ | $10^{15} \Omega^{-1} \cdot m^{-2}$ |
| $G_i$ | $7 \times 10^{13} \Omega^{-1} \cdot m^{-2}$ | $2.4 \times 10^{14} \Omega^{-1} \cdot m^{-2}$ | $4.8 \times 10^{13} \Omega^{-1} \cdot m^{-2}$ |
| $\alpha_R$ | $5 \times 10^{-12} \, eV \cdot m$ | $5 \times 10^{-12} \, eV \cdot m$ | $< 2 \times 10^{-12} \, eV \cdot m$ |

Supplemental Material

**Sec. A. Bi$_2$Te$_3$ film growth**

The Bi$_2$Te$_3$ films were deposited on single-side-polished Al$_2$O$_3$ (0001) substrates by molecular beam epitaxy (MBE). The substrates were cleaned in an ultrasonic bath using acetone and isopropanol, then followed by de-ionized water rinsing. After loading into the MBE growth chamber, the substrates were annealed at 800 °C for 30 minutes in the vacuum with a base pressure below 1.3×10$^{-9}$ Torr to improve the surface quality. A Bi$_2$Te$_3$ film was grown by the evaporation of elemental Bi (6N) and Te (5N) solid sources from two standard Knudsen cells. A two-step deposition method was used to reduce Te vacancies in Bi$_2$Te$_3$ films. Initial 1–2 quintuple layers of Bi$_2$Te$_3$ were deposited at a fixed substrate temperature of 120 °C. Then the substrate temperature was increased to 200 °C under Te flux, and the second step deposition was conducted for the rest quintuple layers of Bi$_2$Te$_3$. Te/Bi beam flux ratio was kept around 25 throughout the whole deposition process. After deposition, the bare Bi$_2$Te$_3$ films were immediately transferred into a magnetron sputtering chamber within 5 mins.

**Sec. B. Fabrication of ST-FMR and spin pumping devices**

The Bi$_2$Te$_3$/Py ST-FMR devices in this work were prepared through sputtering, photolithography, Ar ion milling and lift-off processes. First, a 6-nm Py thin film and 5-nm SiO$_2$ capping layer were deposited on top of MBE-grown Bi$_2$Te$_3$ thin films using magnetron sputtering with a base pressure < 2×10$^{-9}$ Torr. The deposition rate of Py was kept less than 1 nm/min. Then, the Bi$_2$Te$_3$/Py bilayer was patterned into strips with a length of 10–25 µm and width of 15–30 µm as current channels. Next, a Ta (2 nm)/Cu (150 nm)/Pt (3 nm)-structured waveguide was fabricated. For Bi$_2$Te$_3$/Py spin pumping devices, the Bi$_2$Te$_3$/Py/SiO$_2$ stack is the same as ST-FMR.

Then, Bi$_2$Te$_3$/Py bilayer strips were fabricated using photolithography and Ar ion milling. As a next step, 35-nm-SiO$_2$ layer with a size of 640 μm × 640 μm was fabricated on top of the strips to insulate the device and the waveguide, followed by the fabrication of the Ta (2 nm)/Cu (150 nm)/Pt (3 nm)-structured waveguide and two electrodes.

### Sec. C. ST-FMR analysis

The ST-FMR signal measured at each frequency is superimposed by a symmetric and antisymmetric Lorentzian component, which can be decomposed by fitting $V_{\text{mix}}$ to

$$V_{\text{mix}} = V_S \frac{\Delta^2}{\Delta^2+(B-B_0)^2} + V_A \frac{\Delta(B-B_0)}{\Delta^2+(B-B_0)^2}, \tag{1}$$

where $V_S$ and $V_A$ are the amplitudes of the symmetric and antisymmetric Lorentzian component, respectively. $V_S$ and $V_A$ are related to $\tau_\parallel$ and $\tau_\perp$ by $V_S = -\frac{I_{\text{rf}}}{2}\frac{dR}{d\theta_B}\frac{1}{\alpha\gamma(2B_0+\mu_0 M_{eff})}\tau_\parallel$ and $V_A = -\frac{I_{\text{rf}}}{2}\frac{dR}{d\theta_B}\frac{\sqrt{1+\mu_0 M_{\text{eff}}/B_0}}{\alpha\gamma(2B_0+\mu_0 M_{\text{eff}})}\tau_\perp$ [1,2], where $I_{\text{rf}}$ is the r.f. current flowing through the device and $dR/d\theta_B$ is the angular-dependent magnetoresistance at $\theta_B = 40°$, which is measured separately on the same device.

The charge-to-spin conversion efficiency $\xi_{\text{C-S}}$ that characterizes the strength of the in-plane spin orbit torque per unit applied current density at $\theta_B = 0°$ is given by $(2e/\hbar)\sigma_s/\sigma$ [1], where $\sigma$ and $\sigma_s$ are the charge conductivity and the spin conductivity in Bi$_2$Te$_3$ (Pt), respectively. $\sigma_s$ is defined as the in-plane spin-polarized current density per unit electric field, i.e., $\sigma_s = J_s/E = \tau_\parallel M_S t_{\text{FM}}/(E\cos\theta_B)$, where $J_s$ is the in-plane spin-polarized current density absorbed by the ferromagnet at $\theta_B = 0°$. $E$ is the electric field across the device.

## Sec. D. Spin pumping analysis

The DC spin current generated by spin pumping is expressed as [3]

$$j_S = \frac{\omega}{2\pi} \int_0^{2\pi/\omega} \frac{\hbar}{4\pi} g_{eff}^{\uparrow\downarrow} \frac{1}{M_s^2} \left[ M(t) \times \frac{dM(t)}{dt} \right]_z dt, \qquad (2)$$

where $g_{eff}^{\uparrow\downarrow}$ is the real part of spin mixing conductance and $\omega$ is the ferromagnetic resonance frequency. $M_s$ is the saturation magnetization of the ferromagnetic (FM) layer. In a simplified assumption where $m$ is subjected to a circular precession trajectory [4,5], the spin pumping induced DC spin current is quantified by [3]

$$J_S^0 = \frac{2e}{\hbar} \frac{\hbar g_{eff}^{\uparrow\downarrow} \gamma^2 h_{rf}^2 (M_s \gamma + \sqrt{M_s^2 \gamma^2 + 4\omega^2})}{8\pi\alpha^2 (M_s^2 \gamma^2 + 4\omega^2)}, \qquad (3)$$

where $h_{rf}$ is the r.f. magnetic field generated by the coplanar waveguide and calculated from Ampere's law based on $2W_{SG} h_{rf} = \sqrt{\frac{2P_{inp}}{Z}}$, where $W_{SG}$ is the width of the signal line with a value of 60 μm, the input power $P_{inp}$ is 15 dBm, and $Z$ is the impedance of the waveguide characterized by the vector network analyzer. The spin current injected to the non-magnetic (NM) layer decays due to the spin relaxation and diffusion, such that the spin current at distance $y$ from the NM/FM interface is [5]

$$J_S(y) = J_S^0 \frac{\sinh[(d_{NM}-y)/\lambda_{NM}]}{\sinh(d_{NM}/\lambda_{NM})}, \qquad (4)$$

where $d_{NM}$ and $\lambda_{NM}$ are the thickness and spin diffusion length of the Bi$_2$Te$_3$ (Pt) layer respectively, and $y$ is the distance between a position at the NM layer to the NM/FM interface. Therefore, by considering the spin diffusion in the NM layer, the charge current density converted from the spin current can be expressed by

$$\frac{V_S}{R} = (\xi_{s-c} w d_{NM}) \zeta \frac{2e}{\hbar} \frac{\hbar g^{\uparrow\downarrow} \gamma^2 h_{rf}^2 (M\gamma + \sqrt{M^2\gamma^2 + 4\omega^2})}{8\pi\alpha^2 (M^2\gamma^2 + 4\omega^2)} \times \frac{\lambda_{NM}}{d_{NM}} \tanh(\frac{d_{NM}}{2\lambda_{NM}}), \qquad (5)$$

where $w$ is the width of the device and $\zeta$ is a scale factor in order to take into account the fact that only a part of the sample contribute to the spin pumping signal [6]. The width of signal line and ground line of the coplanar waveguide are 60 μm and 180 μm, respectively. Therefore, the r.f. field below the ground line is 3 times smaller and the effective length of device contributing to spin pumping is 120 μm, while the length of device is 800 μm. Therefore, $\zeta$ is determined to be 0.15 (120/800 = 0.15). Here, we assume the spin diffusion length in Bi$_2$Te$_3$ and Pt are to be ~1.0 nm and ~3.4 nm, respectively, based on previous reports [7-9].

**Sec. E. Discussion on the artifact signals in ST-FMR and SP measurements**

**SP signal involved in ST-FMR measurements.** The spin pumping contribution $V_{sp}$ to the total voltage $V_{sym}$ in our samples has been evaluated. $V_{sp}$ takes up to 0.14% and 0.004% in the total $V_{sym}$ for Bi$_2$Te$_3$/Py and Pt/Py, respectively. The inverse spin Hall voltage in the spin pumping process is given by [1,8]

$$V_{sp} = \theta_m \frac{eW\lambda_S R}{2\pi} \tanh(\frac{t}{2\lambda_S}) \text{Re}(g_{\uparrow\downarrow}^{\text{eff}}) \omega \phi_P^2 \sin(\theta_H) \sqrt{H_{\text{ext}}/(H_{\text{ext}} + 4\pi M_{\text{eff}})}, \qquad (6)$$

where $V_{sp}$ is the spin pumping signal in the ST-FMR measurements, $W$ is the microstrip width, $R$ is the device resistance, $\lambda_S$ is the spin diffusion length in Bi$_2$Te$_3$, here we assume it is 1 nm [8]. $\theta_m$ is the measured spin Hall angle (same with the SOT efficiency), $t$ is the Bi$_2$Te$_3$ thickness, $\text{Re}(g_{\uparrow\downarrow}^{\text{eff}})$ is the real part of the effective spin mixing conductance (~7.22×10$^{19}$ m$^{-2}$ for Bi$_2$Te$_3$/Py and ~3.43×10$^{19}$ m$^{-2}$ for Pt/Py, determined from both ST-FMR and SP measurements, the values are consistent with both measurements), and $\theta_H$ is the angle between $I_{RF}$ and magnetic field $H_{\text{ext}}$, which is 40°. $I_{rf}$ is the RF current flowing through the device. $\phi_P$ is the maximum precession angle in the device plane, given by $\phi_P = \frac{1}{dR/d\theta_H} \frac{2}{I_{rf}} \sqrt{S^2 + A^2}$, where $S$ and $A$ are the symmetric and antisymmetric components of the ST-FMR signal, respectively. $\phi_P$ is calculated to be 0.013 *rad* and 0.054 *rad* for Bi$_2$Te$_3$/Py and Pt/Py, respectively. To estimate an upper bound in $V_{sp}$, it is noted that the value of $\lambda_S \, tanh(\, t/2\lambda_S)$ is always less than $t/2$, so we use this as an upper limit. The upper limit ratios of $V_{sp}/S$ are thus calculated to be 0.14% and 0.004% for Bi$_2$Te$_3$/Py and Pt/Py, respectively. Hence, we confirm that the spin pumping contributions in the ST-FMR measurements based on Bi$_2$Te$_3$/Py and Pt/Py are negligible.

**Spin rectification signal involved in SP measurements.** Based on our SP measurement configuration, where the $H_{ext}$ is perpendicular to the voltage leads across the ferromagnet/nonmagnet bilayer, the anisotropic magnetoresistance induced spin rectification signal (AMR-SRE) only contribute to the antisymmetric component of the measured signal. Therefore, AMR-SRE does not influence our analysis results which are only based on the symmetric component [5,10,11].

Besides AMR-SRE, the planar Hall effect (PHE) can also contaminate the SP induced ISHE signal. It is important to note that the spin pumping induced ISHE signal has only symmetric Lorentzian line shape ($V_S$), and must change sign with reversing Py magnetization irrespective of the microwave frequency. Meanwhile, PHE-SRE signal is odd under $H_{ext}$ (Py magnetization) reversal. If PHE-SRE signal plays a significant role, the measured $V_S$ of SP signal should have different amplitude and even same sign under $H_{ext}$ reversal [11,12]. However, in our measurements, as shown in Fig. 2(e) in the main text, the amplitudes of $V_S$ before and after $H_{ext}$ reversal are always the same at different microwave frequencies. More than 12 devices with various sizes are measured for both Bi$_2$Te$_3$/Py and Pt/Py SP samples, we did not observe any measurable $V_S$ and $V_a$ amplitude changes when reversing $H_{ext}$. Therefore, the SP and ST-FMR analyses in this work are not contaminated by both AMR-SRE and PHE-SRE.

**Angular-dependent ST-FMR measurements.** As suggested by the reviewer, the angular dependent ST-FMR measurements have been performed on $Bi_2Te_3$/Py and Pt/Py samples. As shown in Fig. S1, $V_S$ and $V_A$ can be well fitted by $A\sin2\varphi\cos\varphi$ for both samples, where $\varphi$ is the angle between the magnetization and current direction, suggesting that there is no measurable unidirectional spin-torque driven magnetization dynamic contribution in our ST-FMR measurements [13].

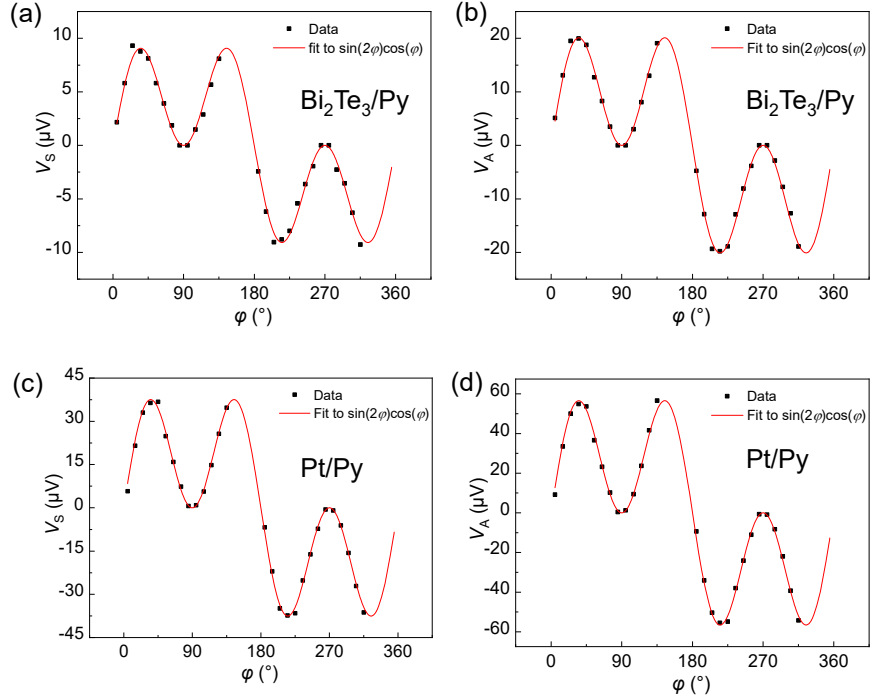

**Fig. S1** Angular dependence of ST-FMR signals. (a-d) Symmetric ($V_S$) and antisymmetric ($V_A$) ST-FMR resonance components for $Bi_2Te_3$ (13 nm)/Py (6 nm) and Pt (5 nm)/Py (6 nm) samples as a function of in-plane magnetic field angle. The microwave frequency is 7 GHz and the applied microwave power is 15 dBm.

**References for Supplemental Material**